\documentclass[%
aip,amsmath,amssymb,reprint]{revtex4-2}

	\usepackage{bm}
	\usepackage[utf8]{inputenc}
	\usepackage[T1]{fontenc}
	\usepackage{mathptmx}	
	\usepackage{microtype}

	\usepackage{tikz,pgf,placeins}
	\graphicspath{{img/}}

	\usepackage{amsfonts}
	\usepackage{amsmath}
	\usepackage{amssymb}
	\usepackage{dcolumn}
	\usepackage{enumerate}
	\usepackage[version=4]{mhchem}
	\usepackage{siunitx}
		\DeclareSIUnit\sccm{sccm}
		\DeclareSIUnit\torr{torr}
	\usepackage{soul}
	\usepackage{xcolor}
	
	\newcommand*{\B}[1]{\relax\ifmmode\bm{#1}\else\textbf{#1}\fi}
	\newcommand*{\Tc}{\ensuremath{T_\text{c}}}
	\newcommand*{\kB}{\ensuremath{k_\text{B}}}
	\newcommand*{\Sum}{\displaystyle\sum}

	\usepackage{graphicx}

	\usepackage[colorlinks,allcolors=blue]{hyperref}
	\usepackage[capitalize]{cleveref} 
	
	\definecolor{Blue}{RGB}{0,0,120}

	\definecolor{Green}{RGB}{0,120,0}

	\definecolor{Purple}{RGB}{220,0,220}
	
	\definecolor{Red}{RGB}{220,0,0}

\begin{document}
\title[Influence of substrate-induced thermal stress on the superconducting properties of \ce{V3Si} thin films]{Influence of substrate-induced thermal stress on the superconducting properties of \ce{V3Si} thin films}

\author{T.D. Vethaak}
\affiliation{Université Grenoble Alpes, CEA, IRIG-Pheliqs, 38000 Grenoble, France}
\affiliation{CEA-LETI-Minatec, 17 rue des Martyrs, 38054 Grenoble, France}
\author{F. Gustavo}
\affiliation{Université Grenoble Alpes, CEA, IRIG-Pheliqs, 38000 Grenoble, France}
\author{T. Farjot}
\affiliation{CEA-LETI-Minatec, 17 rue des Martyrs, 38054 Grenoble, France}
\author{T. Kubart}
\affiliation{Division of Solid-State Electronics, Department of Electrical Engineering, Uppsala University, 75121 Uppsala, Sweden.}
\author{P. Gergaud}
\affiliation{CEA-LETI-Minatec, 17 rue des Martyrs, 38054 Grenoble, France}
\author{S-L. Zhang}
\affiliation{Division of Solid-State Electronics, Department of Electrical Engineering, Uppsala University, 75121 Uppsala, Sweden.}
\author{F. Nemouchi}
\affiliation{CEA-LETI-Minatec, 17 rue des Martyrs, 38054 Grenoble, France}
\author{F. Lefloch}
\affiliation{Université Grenoble Alpes, CEA, IRIG-Pheliqs, 38000 Grenoble, France}

\date\today

\begin{abstract}
	\noindent Thin films of superconducting \ce{V3Si} were prepared by means of RF sputtering from a compound \ce{V3Si} target at room temperature onto sapphire and oxide-coated silicon wafers, followed by rapid thermal processing under secondary vacuum.
	The superconducting properties of the films thus produced are found to improve with annealing temperature, which is ascribed to a reduction of defects in the polycrystalline layer.
	Critical temperatures (\Tc) up to \SI{15.3}{\kelvin} were demonstrated after thermal processing, compared to less than \SI{1}{\kelvin} after deposition.
	The \Tc\ was found to always be lower on the silicon wafers, by on average \SI{1.9\pm0.3}{\kelvin} for the annealed samples.
	This difference, as well as a broadening of the superconducting transitions, is nearly independent of the annealing conditions.
	In-situ XRD measurements reveal that the silicide layer becomes strained upon heating due to a mismatch between the thermal expansion of the substrate and that of \ce{V3Si}.
	Taking into account the volume reduction due to crystallization, this mismatch is initially larger on sapphire, though stress relaxation allows the silicide layer to be in a relatively unstrained state after cooling.
	On oxidized silicon however, no clear evidence of relaxation upon cooling is observed, and the \ce{V3Si} ends up with an out-of-plane strain of 0.3\% at room temperature.
	This strain increases as the sample is cooled down to cryogenic temperatures, though the deformation of the polycrystalline layer is expected to be highly inhomogeneous.
	Taking into account also the reported occurrence of a Martensitic transition just above the critical temperature, this extrapolated strain distribution is found to closely match an existing model of the strain dependence of A-15 superconducting compounds.
\end{abstract}

\maketitle

\section{Introduction}
	Self-aligned silicides have long been the technology of choice for contacting CMOS transistors~\cite{shibata1981optimally,zhang2014metal}, as they offer low specific resistance, a reliable process, and often a silicidation-tunable Schottky barrier height~\cite{nishi2011schottky}.
	Though recent efforts have focused mainly on nickel and nickel-platinum silicides, numerous other transition metals have been studied for this application, each often presenting its own advantages.
	
	As the downscaling progressed and requirements shifted, these specific advantages have often meant that material preferences shifted, most recently as Ni(Pt)Si superseded \ce{CoSi2}~\cite{lavoie2003towards,zhang2003metal}.
	Now, with the end of Moore's law in sight and interest in beyond-CMOS technologies rising~\cite{waldrop2016chips,rupp2010economic}, such demands may shift once more.
	One area of investigation is superconducting computing, which offers lower power requirements for large-scale operation and, most prominently, provides a platform for quantum computing.
	The central element of such technologies is the Josephson junction~\cite{josephson1962possible,josephson1964coupled}, traditionally made by joining two superconductors with an insulating barrier, across which superconducting carriers can tunnel.
	Because transport through such insulators cannot be modulated by applying a voltage, gates typically rely on tuning the interference between two nearby junctions with magnetic fields~\cite{devoret2005implementing}.
	Such magnetic tuning has the disadvantages that the fields from nearby gates may affect each other, and that it effectively doubles the number of junctions that need to be patterned.
	A different approach returns to the mature CMOS process, drawing on existing designs and fabrication expertise to create Josephson field effect transistors (JoFETs)~\cite{clark1980feasibility}.
	In such devices, superconducting silicides offer high-quality contacts to a semiconducting channel, allowing the superconducting transport to be tuned by a gate voltage.
	In addition to most of the usual material properties desired of silicides for regular transistors, superconducting behavior is now of central importance.
	
	Not all silicides are superconducting, and those that are have widely different critical temperatures ($\Tc$'s) at which superconductivity appears.
	PtSi and \ce{CoSi2} for example, which are relatively mature in terms of integration, offer values of around 1.0 and \SI{1.4}{\kelvin}~\cite{prest2015superconducting,berger1997properties}, whereas the less developed \ce{V3Si} can be superconducting up to \SI{17}{\kelvin}~\cite{hardy1953superconducting,blumberg1960correlations}.
	A high \Tc\ is desirable as it relaxes the cryogenic requirements for operation, but it also directly relates to the strength with which superconducting electrons bind to each other.
	This can be quantified by the energy gap $\Delta$ (not to be confused with the semiconducting band gap), which in general relates to the critical temperature through $\Delta\propto\kB\Tc$.
	To illustrate, while PtSi has a superconducting band gap of 70--\SI{150}{\micro\electronvolt} depending on the film thickness~\cite{prest2015superconducting}, that of \ce{V3Si} is in the range of $2.2$--$\SI{2.8}{\milli\electronvolt}$~\cite{moore1979energy,hauser1966energy}.
	A higher critical temperature therefore means that the electron pairs are harder to break apart as they traverse the semiconducting channel, and coherent transport between the contacts is easier to establish.
	It also implies a shorter coherence length, around 7~nm for \ce{V3Si}~\cite{meyer1969experimentelle}, making the superconducting behavior near an interface more sensitive to the quality of the surface, and thus imposing more stringent requirements on the fabrication~\cite{schumann1979tunneling}.
	Besides the critical temperature, other properties such as the Schottky barrier height to silicon, mismatches with the channel in effective mass and lattice parameter, and the ability to reliably form high-quality films are important in selecting a silicide for Josephson field effect transistors.
	
	Though \ce{V3Si} offers the highest known critical temperature of any silicide that the authors are aware of, relatively little is known about its integration into CMOS devices.
	Motivated by this scarcity in the literature, we aim to understand the formation of thin films and their behavior at cryogenic temperatures.
	The focus in this study is the apparent influence of the substrate type on the superconducting properties of the thin film, which seems not to be related to the qualities of the film itself.
	Analysis of changes in the the out-of-plane lattice parameter induced during and after thermal processing suggests that the residual strain is substrate-dependent, and strongly affects the critical temperature.

\section{\ce{V3Si} deposition and thermal processing}

	Among vanadium silicides predicted by the V/Si phase diagram, only \ce{VSi2} and \ce{V3Si} form in thin-film conditions (few nm to µm).
	Since the reaction of pure vanadium and silicon usually leads to the undesired compound \ce{VSi2}~\cite{tu1973formation}, it is necessary to modify the conventional self-aligned process of metal deposition followed by a combination of thermal processing and selective etch.
	Four approaches were considered to select the \ce{V3Si} phase: (i) deposition of pure vanadium on \ce{SiO2}~\cite{krautle1973silicide}, (ii) deposition of V on O-doped Si~\cite{schutz1979formation}, (iii) sequential sputtering of V and Si in the right stoichiometry~\cite{de1985preparation}, and (iv) direct sputtering of \ce{V3Si} from a compound target~\cite{michikami1982v3si}.
	The first two rely on the presence of oxygen to prevent the \ce{VSi2} phase, possibly by slowing the diffusion of Si and so selecting V as the dominant diffusing species~\cite{schutz1979formation}. 
	At one extreme, V was deposited on relatively thick \ce{SiO2} layers, which would reliably form \ce{V3Si} at the interface, with \ce{VO_x} accumulating on the surface~\cite{krautle1974reactions,krautle1974kinetics}.
	Similar results could be obtained at much lower oxygen concentrations, when the metal was deposited on oxygen-doped amorphous or crystalline silicon~\cite{schutz1979formation}.
	Either technique has the disadvantage of necessitating the selective removal of an oxide layer, and likely leaving residual oxygen contamination in the silicon.
	The latter would affect the contact resistance, and could impair device performance at cryogenic temperatures by providing parasitic two-level systems.
	Instead of adding reaction inhibitors that prevent the formation of \ce{VSi2}, the nucleation of \ce{V3Si} can be aided by providing the right local concentration of the respective elements.
	Both sequential sputtering of Si and V~\cite{de1985preparation} and deposition from a compound \ce{V3Si} target~\cite{theuerer1964getter,michikami1982v3si} have been shown to be effective to this end.
	We have opted for this last method, which can ensure continuity to lower film thicknesses than sequential sputtering, and requires only a single target slot in the deposition tool.
	The clear downside of this process is that while with the salicide process the silicon is only provided by the contact openings connected to the channel, it is now present across the entire surface, and hence the silicidation no longer occurs only locally.
	It is possible to address this issue with appropriate process modifications.

	The \ce{V3Si} target had a measured silicon content of \SI{22.7\pm0.2}{}~at.\% (WDXRF) to \SI{25.7}{}~at.\% (supplier data), which may differ from the \ce{Si} concentration in the deposited layer by \SI{1}{}~at.\%~\cite{theuerer1964getter}.
	Conflicting reports exist on the impact of a deviation from Si:V stoichiometry on the critical temperature: a reduction in Si content from 25 to 20\% has been observed to reduce the critical temperature by \SI{6}{\kelvin}~\cite{junod1971superconductivity}, while others report that a \SI{\pm20}{\percent}at change in Si content \emph{``produced hardly any change of \Tc.''}~\cite{hardy1954superconductivity}.
	It has been argued elsewhere that while the chemical composition itself is unimportant, any changes in the lattice parameter that it induces may cause large reductions in \Tc~\cite{testardi1970unusual}.
	The \ce{V3Si} films were deposited on substrates using RF magnetron sputtering equipment.
	The pressure, Ar flow and RF power were optimized to obtain the desired \ce{V3Si} stoichiometry of the films.
	During  the film deposition, the Ar pressure was maintained at \SI{2.5E-3}{\milli\bar} with Ar flow of \SI{50}{\sccm}.
	The RF sputtering power was \SI{200}{\watt} for all samples.
	Substrates were in rotation and translation movement during deposition.
	The thicknesses of the films were determined by an X-ray reflectivity (XRR) method (D8Fabline).
	
	Though previous investigations on sputtering from a compound \ce{V3Si} target used heated substrates~\cite{theuerer1964getter,michikami1982v3si}, in the current study the chuck was at room temperature (not intentionally heated), so that the heating step could be examined separately.
	The lower deposition temperature led to a relatively higher resistivity of \SI{180\pm10}{\micro\ohm\centi\meter} of the as-deposited layers.
	Subsequent rapid thermal processing (RTP) occurred under secondary vacuum at different temperatures during two minutes in a Jipelec furnace, with a \SI{20}{\celsius/\minute} ramp rate.
	Three types of wafers were used as substrate: silicon with 20~nm of thermal \ce{SiO2}, HF-cleaned silicon, and sapphire. 
	The \ce{V3Si} showed no sign of reacting with the sapphire and \ce{SiO2}-coated silicon wafers during a later annealing, though \ce{VSi2} formation occurred on the cleaned Si substrates.
	In this report we focus on samples where \ce{V3Si} was the only silicide phase present.

\section{\label{sec:low_temperature}Low-temperature measurements}\FloatBarrier

	\begin{figure}
		\centering
		\includegraphics[width=\columnwidth]{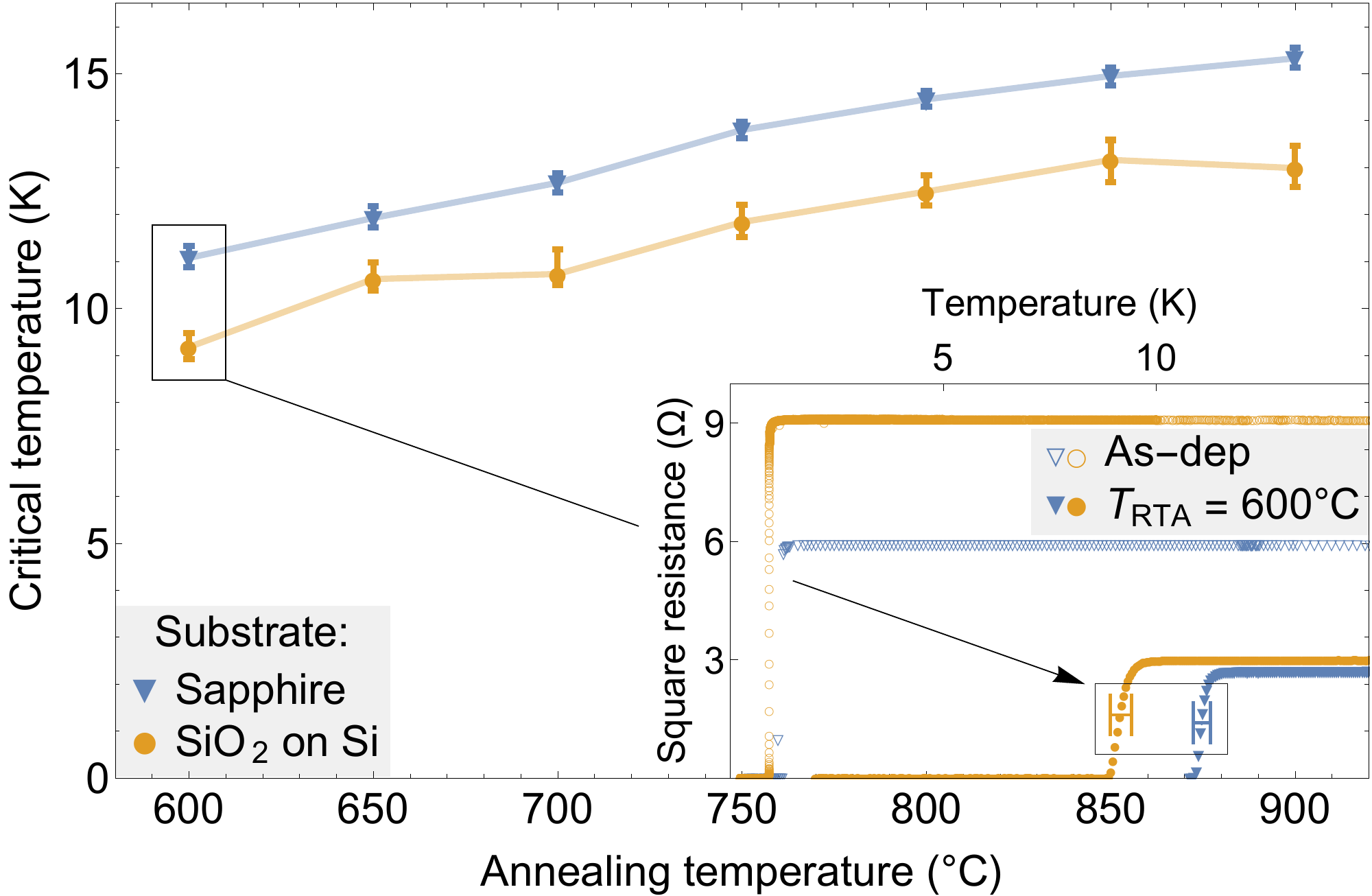}
		\caption{\label{fig:V3SiPlotTcSapphireSiO2}The critical temperature ($T_\text{c}$) of a 200~nm thin film of \ce{V3Si} is plotted versus the temperature at which the sample was annealed, with error bars indicating the temperatures at which the sample retained 10\% and 90\% of the normal-state resistance. \B{Inset:} resistance measurements on selected samples, showing sharp drops to zero as they become superconducting. Measurements on as-deposited samples for both substrates are shown with open symbols.}
	\end{figure}
	
	\begin{figure}
		\centering
		\includegraphics[width=\columnwidth]{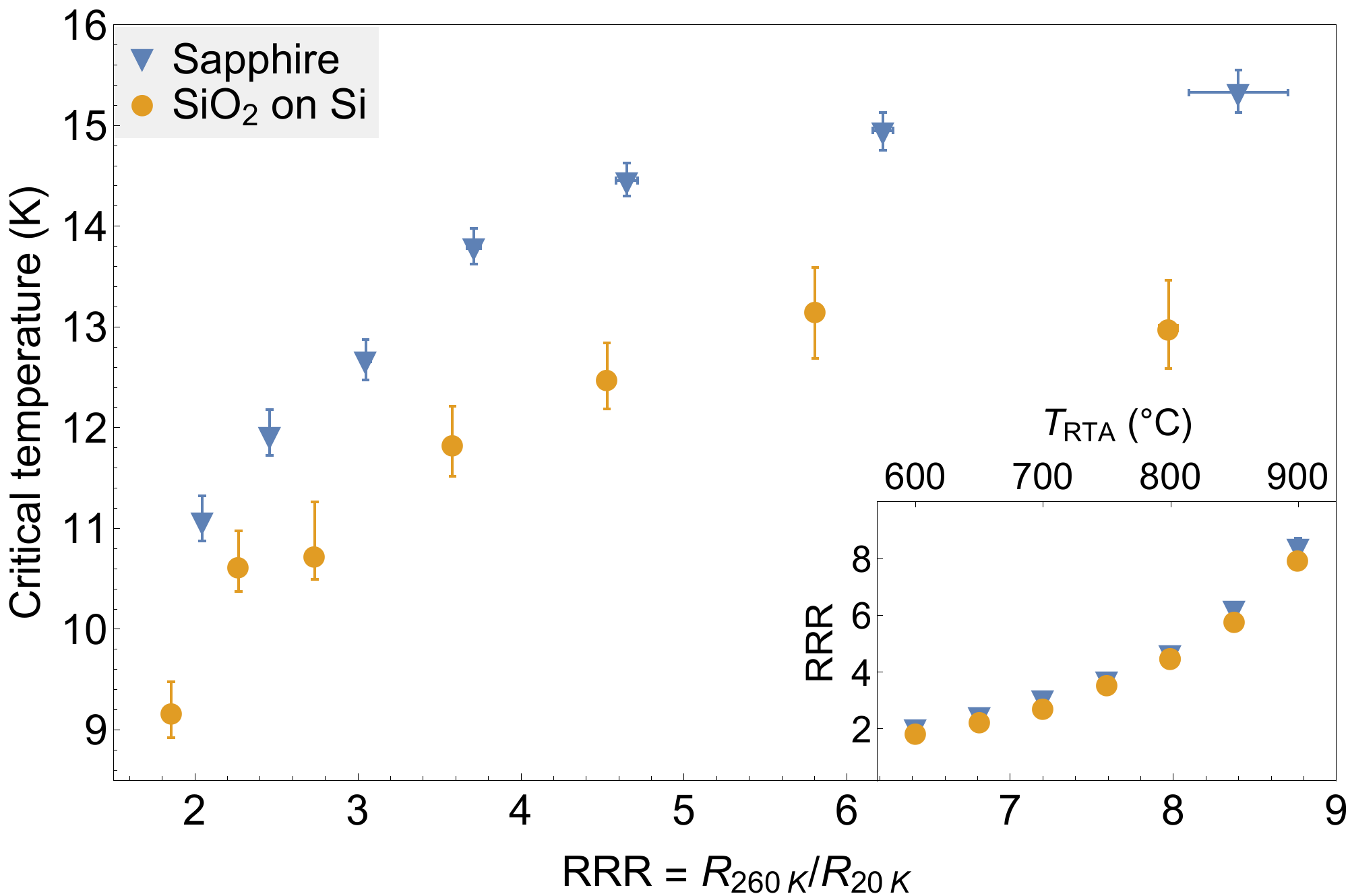}
		\caption{\label{fig:RRRTcErrorPlotSubstratesSapphireSiO2}Residual-resistance ratio (RRR), here defined as the ratio of the resistances at $260$ to that at \SI{20}{\kelvin}, increases with annealing temperature for both substrates, indicating a reduction in defect density or clustering of impurities.}
	\end{figure}
	
	After deposition, the \ce{V3Si} layers were amorphous, with no discernible peaks of the compound present in its XRD diffractogram.
	This is reflected in its superconducting properties, as the critical temperature was only \SI{0.9}{\kelvin} and \SI{1.2}{\kelvin} on the silicon and sapphire substrates, respectively.
	Heating the layers promotes crystallization, and the critical temperature improves markedly, approaching that of bulk samples~\cite{hardy1953superconducting} to within \SI{2}{\kelvin} as the annealing temperature was increased to \SI{900}{\celsius}.
	This is illustrated in the main graph of Fig.~\ref{fig:V3SiPlotTcSapphireSiO2}, where each point represents a sample with \SI{200}{\nano\meter} of \ce{V3Si} that was annealed at the indicated temperature.
	The continuing upward trend at the higher end of the annealing temperature scale suggests that further improvements in \Tc\ could straightforwardly be achieved.
	
	The rise in critical temperature due to annealing can be attributed to the elimination of scattering centers as the material becomes more ordered and the density of defects is reduced.
	This can be quantified by the residual resistance ratio ${\text{RRR}=R_{\SI{260}{\kelvin}}/R_{\SI{20}{\kelvin}}}$, where the reference temperature of \SI{260}{\kelvin} is chosen such that data is available for all samples.
	The electron-electron and electron-phonon interactions that dominate the resistance of a metal at room temperature are strongly suppressed as the material is cooled down.
	The scattering rate on impurities and defects is temperature-independent however, and is therefore the only contribution that remains, hence providing the residual resistance.
	Improvements in superconducting properties are associated with larger resistance ratios~\cite{testardi1977anomalous}, illustrated by the strong relationship in Fig.~\ref{fig:RRRTcErrorPlotSubstratesSapphireSiO2}.
	The inset in this figure shows that there is a clear development of this ratio as the annealing temperature is increased, which is nearly independent of the sample substrate.
	
	When plotted both against the annealing temperature and the residual resistance ratio, a clear offset in \Tc\ of \SI{1.9\pm0.3}{\kelvin} (Fig.~\ref{fig:V3SiPlotTcSapphireSiO2}) and \SI{1.8\pm0.2}{\kelvin} (Fig.~\ref{fig:RRRTcErrorPlotSubstratesSapphireSiO2}) is visible between the samples with Si and sapphire substrates.
	This relative reduction of the critical temperature due to the substrate cannot be explained by different concentrations of impurities or defects, as in this case the points in Fig.~\ref{fig:RRRTcErrorPlotSubstratesSapphireSiO2} would be expected to fall onto a single line.
	Since the sapphire samples were smaller ($10\times\SI{5}{\milli\meter\squared}$, while the Si pieces were $20\times\SI{20}{\milli\meter\squared}$), the modestly higher RRR on some of these pieces could be explained by a faster thermalization during the annealing.
	This cannot, however, account for any of the difference between sapphire and silicon in the relation between RRR and \Tc.
	The substrate therefore appears to affect the superconducting behavior of the \ce{V3Si} without necessarily altering the dynamics of the silicide formation.
	The superconducting transitions on the sapphire substrate are also sharper, with an average transition width $T_\text{90\%}-T_\text{10\%}$ of \SI{3.0\pm0.7}{\percent} of the critical temperature, while that on the silicon substrate is \SI{6.2\pm0.7}{\percent}.
	This suggests that in addition to improving the average superconducting behavior of the material, the sapphire substrate also reduces local variations, causing the material to be more homogeneous at low temperatures.
	
\section{\label{sec:xrd}In-situ XRD analysis}\FloatBarrier
	
	\begin{figure}
		\centering
		\includegraphics[width=\columnwidth]{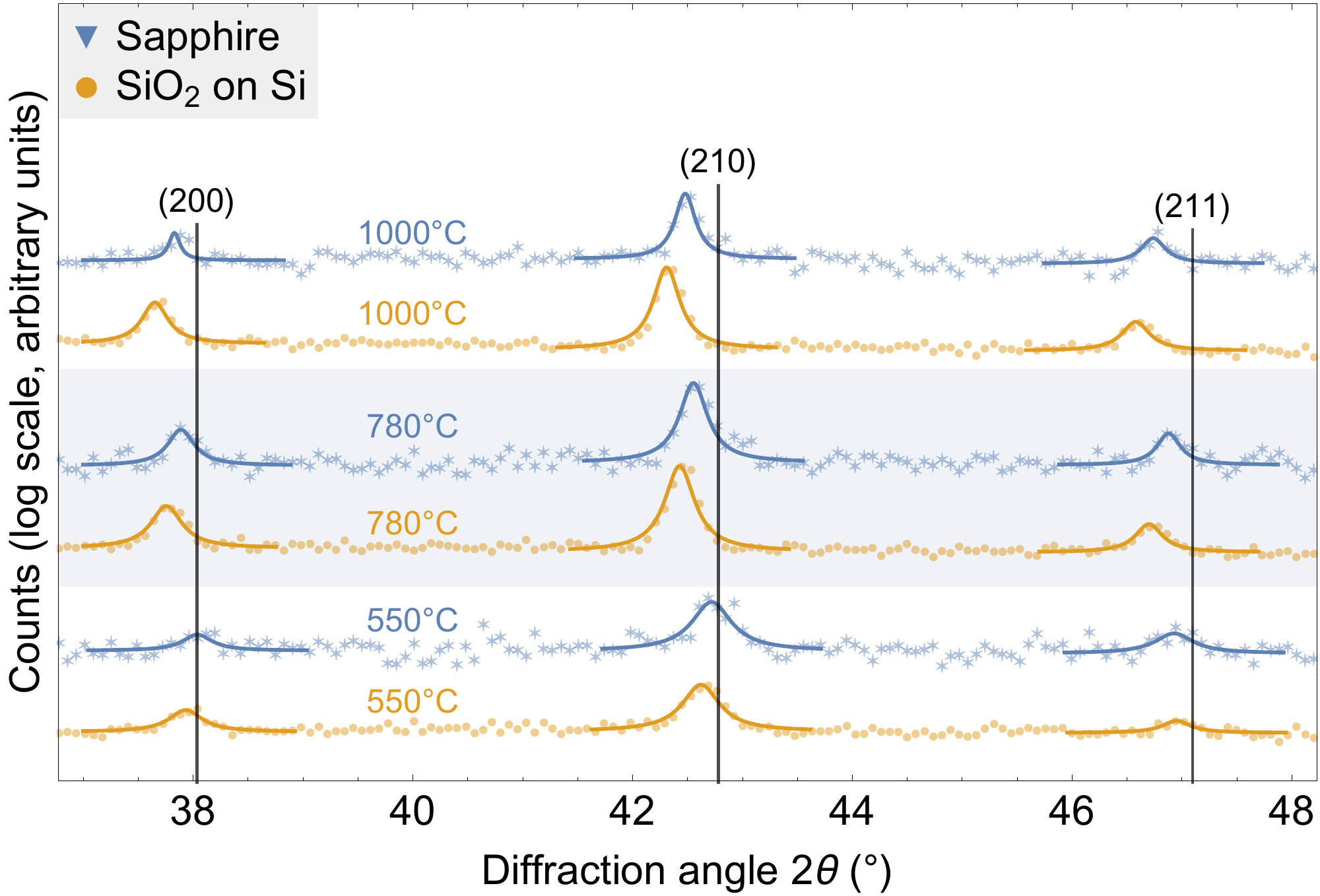}
		\caption{\label{fig:XRD1DPlotOneLine}Selected $\theta$-$2\theta$ curves measured at 550, 780 and \SI{1000}{\celsius}, reference peak positions of the (200), (210) and (211) planes are indicated by black lines.}
	\end{figure}
	
\begin{figure}
		\centering
		\includegraphics[width=\columnwidth]{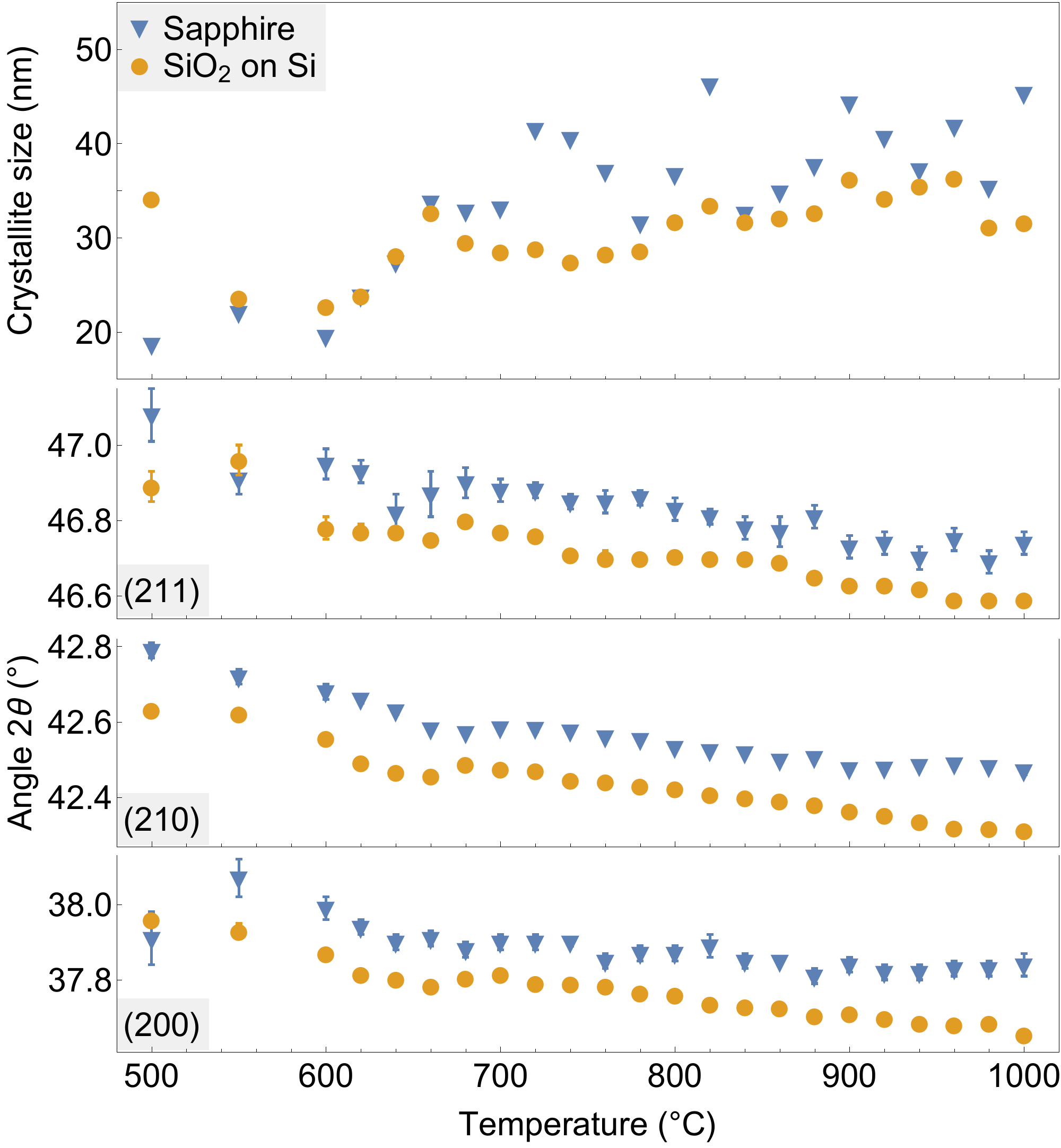}
		\caption{\label{fig:XRDPlot2ThetaAndGrainSize}The grain size and peak positions extracted from in-situ XRD data. An increase in grain size with sample temperature is evident, though no clear dependence on the substrate is observed (a slight divergence at higher temperatures is ascribed to a difference in instrumental line broadening between the two measurements).}
	\end{figure}

	One sample of each substrate with 200~nm of \ce{V3Si} was heated in steps of \SI{50}{\celsius} up to \SI{600}{\celsius} and then in steps of \SI{20}{\celsius} to \SI{1000}{\celsius}.
	The temperature was held constant after each step for around 15~minutes, and an out-of-plane $\theta$-$2\theta$ XRD scan was performed.
	After the samples cooled down to \SI{30}{\celsius}, a final measurement was taken.
	Even when the constituent elements are already intermixed, it is not expected that they would form the silicide phase unless sufficient thermal energy is supplied to stimulate atomic movement and bond formation, and to overcome the barrier posed by the surface energy of a newly formed crystallite.
	This thermal activation of the formation is demonstrated by the absence of any peaks in the scans up to \SI{450}{\celsius}.
	Three peaks of the cubic A15 \ce{V3Si} phase at \ang{38.0} (200), \ang{42.7} (210) and \ang{47.1} (211) become visible on both substrates at \SI{500}{\celsius}, each becoming sharper and moving to smaller angles as the temperature is increased.
	Individual scans at 550, 780 and \SI{1000}{\celsius} are shown in Fig.~\ref{fig:XRD1DPlotOneLine}, while the shift in peak positions with temperature is made more explicit in Fig.~\ref{fig:XRDPlot2ThetaAndGrainSize}.
	The shift of the peaks to smaller angles is explained by the thermal expansion of the \ce{V3Si}, and is affected by strain imposed by the substrate, which expands at a different rate.
	Using the Scherrer equation~\cite{patterson1939scherrer}, it is possible to estimate the crystallite size from the full width at half maximum of each of the three peaks.
	In the top panel of Fig.~\ref{fig:XRDPlot2ThetaAndGrainSize}, a weighted average (weighted by peak area) of these three values is shown for each temperature, showing a continued crystal growth during the measurement.
	
	Though there is no one-to-one correspondence between the crystallinity observed at a given temperature during the in-situ experiment and that obtained after thermal processing for two minutes at that same temperature, the general relationship is assumed to hold.
	The improvements in resistance ratio and critical temperature with increasing RTP temperature as shown in Figs.~\ref{fig:V3SiPlotTcSapphireSiO2} and~\ref{fig:RRRTcErrorPlotSubstratesSapphireSiO2} can thus likely be attributed at least in part to enhanced crystallinity.
	This does not mean, however, that the differences in superconducting properties between the two substrates can also be explained by the quality of the material.
	If the sapphire substrate were to better promote \ce{V3Si} growth, one would expect to see both larger grains and a higher degree of texture.
	We found no sign of either.
	The grain size developed at a similar rate on sapphire and oxidized silicon, and no significant difference between the two substrates was observed in the distribution of crystal orientations.
	The films are preferentially oriented along the <210> direction on both substrates.
	Nor did any of the three analyzed peaks exhibit growth relative to the others while the temperature was increased, as would be expected in the case of epitaxial alignment.
	The differences in superconducting behavior between the two substrates therefore cannot be ascribed to a relative improvement in crystallinity on the sapphire substrate.

\begin{figure}
		\centering
		\includegraphics[width=\columnwidth]{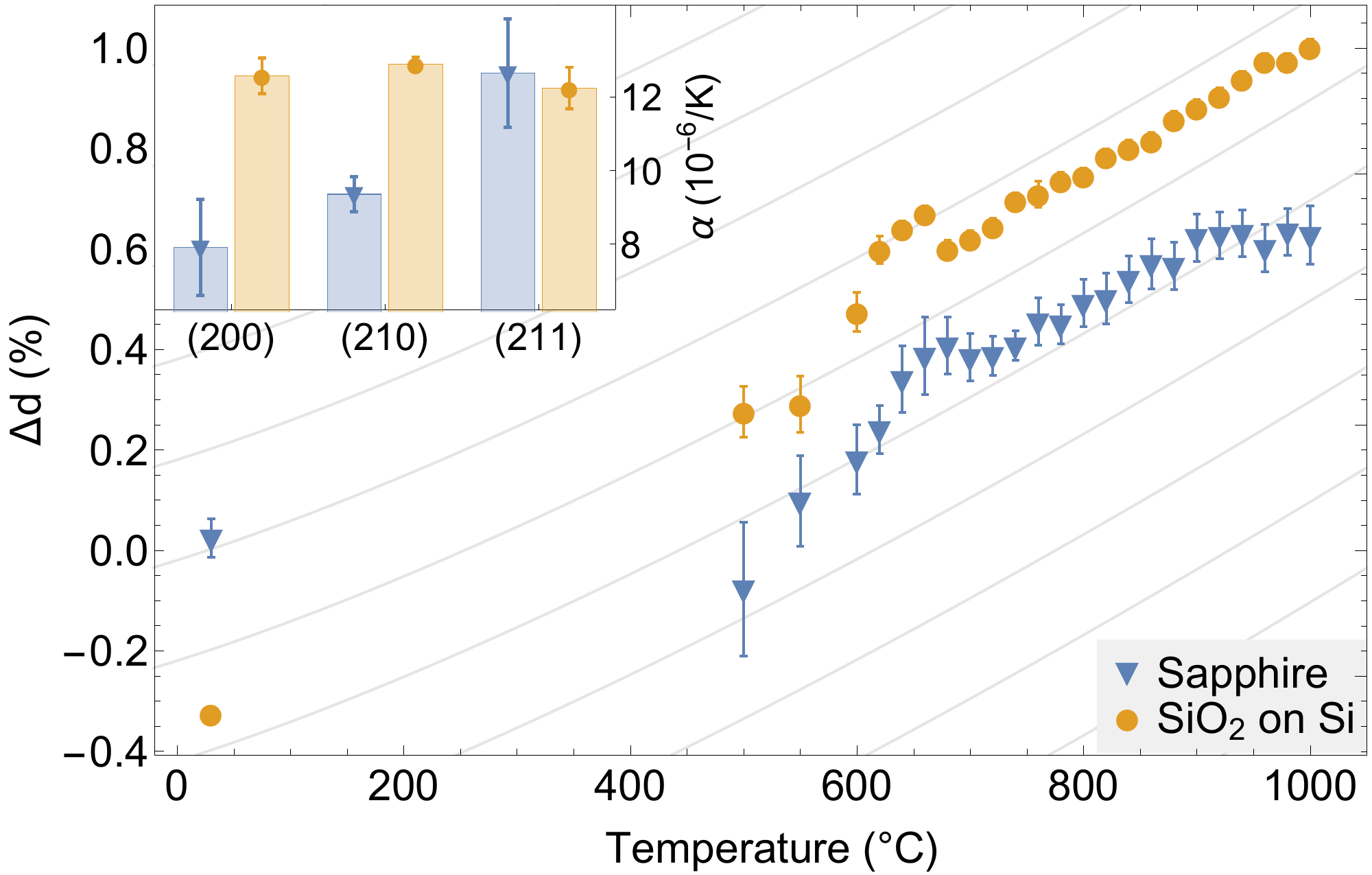}
		\caption{\label{fig:XRDPlotStrainInsetTECPlusRT}The change in out-of-plane lattice parameter with temperature, calculated from the (200), (210) and (211) peak positions. The two points at \SI{30}{\celsius} were recorded at the end of the heating cycle, after the samples cooled back down. Inclined gray lines represent the expected expansion of crystalline bulk \ce{V3Si}, after Ref.~\onlinecite{testardi1972structural}.}
	\end{figure}

	The dependence on temperature of the \ce{V3Si} peak positions was used to calculate the out-of-plane expansion of the crystals.
	In Fig.~\ref{fig:XRDPlotStrainInsetTECPlusRT}, the change in lattice parameter relative to a PDF reference value is shown.
	The parameter was calculated from each of the three peaks, after which an average value was taken.
	This average was weighted by the relative peak areas to account for the larger uncertainty in the position of the smaller peaks.
	An offset between the expansions on silicon and sapphire is immediately apparent, as well as a sudden reduction in volume on both substrates between 660 and \SI{700}{\celsius}.
	The latter is attributed to a plastic deformation of the crystal as strain is relaxed that was built up during the transition from the initial amorphous phase to the more efficiently packed crystal phase.
	
	On the sapphire substrate the rate at which grains expand out of plane is found to depend on their orientation: the grains that are tilted with respect to the horizontal substrate, represented by the (210) and (211) peaks, expand at a higher rate.
	On silicon, however, each peak gives the same expansion rate.
	This is illustrated by the bar graph inset in Fig.~\ref{fig:XRDPlotStrainInsetTECPlusRT}, where each bar represents the average linear thermal expansion coefficient of grains oriented along the indicated Miller indices, calculated with a linear regression on peak positions obtained between 700 and \SI{1000}{\celsius}.
	The fact that the expansion rate depends on the orientation on the sapphire substrate, while it does not on silicon, can be explained by sapphire's larger thermal expansion coefficient.
	When taking into account the volume reduction due to the crystallization of the initially amorphous layer, the \ce{V3Si} likely has an effective expansion rate similar to that of the silicon substrate, so that little thermal stress will occur.
	The situation is different on sapphire, which expands faster and so imposes an in-plane	tensile strain on the silicide layer.
	Sapphire has a room-temperature thermal expansion coefficient of \SI{5.0E-6}{\per\kelvin} orthogonal to the c-axis~\cite{dobrovinskaya2009propertiesp109}, which is the in-plane direction of our substrates, while that of silicon is \SI{2.6E-6}{\per\kelvin} in all directions~\cite{watanabe2004linear}.
	Meanwhile, monocrystalline \ce{V3Si} has an expansion rate of \SI{7.5E-6}{\per\kelvin} at room temperature~\cite{testardi1972structural}.

	Crystals with their main lattice planes aligned normal to the substrate can minimize the change to their unit cell volume by compensating for the in-plane expansion with a reduction in the out-of-plane lattice parameter.
	As a result, the (200) peaks in the $\theta$-$2\theta$ scans exhibit a smaller thermal expansion coefficient on sapphire than on silicon.
	When crystals are tilted with respect to the imposed strain, however, 
	they will either undergo a shearing deformation or see a change in their unit cell volume, both of which are energetically unfavorable.
	The more misaligned a grain is, the harder it therefore is to compensate for in-plane tensile strain by out-of-plane contraction, which explains why those oriented along the <211> axis have a larger out-of-plane thermal expansion coefficient than the <210> grains, which in turn expand
	faster than those oriented <200>.
	
	Two more XRD scans were performed after the samples were cooled down to room temperature, visible on the bottom-left of Fig.~\ref{fig:XRDPlotStrainInsetTECPlusRT}.
	Residual strain is present in the \ce{V3Si} on the silicon substrate, as the out-of-plane lattice parameter is -0.3\% smaller than expected for an unstrained crystal.
	This is likely caused by an in-plane tensile stress imposed during the cooling by the relatively small shrinking of the substrate, which translates to an out-of-plane compressive strain as volume change is minimized.
	No such residual strain is observed on the sapphire substrate, where the silicide's room-temperature lattice parameter closely matches that of the reference value.
	A contraction of the \ce{V3Si} during cooling from \SI{1000}{\celsius} to room temperature, unrestricted by a substrate, would follow the isobaric gray lines indicated in Fig.~\ref{fig:XRDPlotStrainInsetTECPlusRT}.
	Though sapphire's thermal expansion coefficient is larger than that of silicon, it is still smaller than that of crystalline \ce{V3Si}, so also on this substrate one would expect thermal stress to further compress the out-of-plane lattice parameter.
	If no stress relaxation were to occur on sapphire, we would therefore expect the \ce{V3Si} lattice parameter to be reduced by more than \SI{-0.5}{\percent} out-of-plane at room temperature.
	The fact that no residual strain is observed on the sapphire substrate thus suggests that the induced stresses, large compared to those present on the oxidized silicon, have been relaxed by a plastic deformation during the cooling.
	
\section{Discussion}
	
	The superconducting properties of thin films of \ce{V3Si} prepared by means of RF sputtering from a compound \ce{V3Si} target at room temperature were found to be comparable to or better than those of silicide films formed after pure metal deposition~\cite{oya1982superconducting,oya1984enhanced,schutz1979formation} or by molecular-beam epitaxy~\cite{croke1988growth}.
	Separating the deposition and thermal processing steps also allowed improvements relative to hot deposition from a compound target~\cite{theuerer1964getter,michikami1982v3si}.
	
	The dependence of the superconducting properties of the films on the substrate type can be attributed to thermal stress.
	Testardi and coworkers developed a thermodynamic model in the 1970's that describes the general dependence of the critical temperature of A-15 superconductors on strain~\cite{testardi1970unusual,testardi1971unusual95,testardi1971unusual}.
	A second-order expansion in strain coefficients $\epsilon_i$ was proposed, where the coefficients were determined using sound velocity and specific heat measurements.
	Given a $6\times1$ strain matrix $\epsilon$, this dependence was given as
	\begin{equation}\label{eq:testardi_eq2}
		\Tc(\epsilon)-\Tc(0)=
		\Sum_i\Gamma_i\epsilon_i+\dfrac{1}{2}\Sum_i\Sum_j\Delta_{ij}\epsilon_i\epsilon_j.\end{equation}
	In the case of \ce{V3Si} it was found that ${\Delta_{11}=\SI{-24E4}{\kelvin}}$, ${\Delta_{12}=\SI{-5E4}{\kelvin}}$, ${\Delta_{44}=\SI{-1E4}{\kelvin}}$, and ${|\Gamma|<\SI{50}{\kelvin}}$, with all other $\Delta_{ij}=0$.
	
	If the volume remains constant, as it is reported to do precisely~\cite{batterman1966low}, then equal strains $\epsilon_2=\epsilon_3$ applied along the two in-plane axes are associated with an out-of-plane strain $\epsilon_1$ of twice their magnitude (assuming $\epsilon_i\ll1$):
	\begin{equation}\epsilon=\left(\epsilon_1,-\dfrac{\epsilon_1}{2},-\dfrac{\epsilon_1}{2},0,0,0\right),\;\;\;\text{with}\;\;\;\epsilon_1=\dfrac{2}{3}\left(\dfrac{c}{a}-1\right).\end{equation}	
	This means that only the terms with $i,j=1,2,3$ need to be kept in eq.~\eqref{eq:testardi_eq2}, which simplifies to
	\begin{equation}\label{eq:testardi_simplified}\Tc(\epsilon)-\Tc(0)= \dfrac{3}{4}\,\epsilon_1^2\left(\Delta_{11}-\Delta_{12}\right).\end{equation}
	The coefficients $\Delta_{11}$ and $\Delta_{12}$ in this expression were originally determined for bulk samples, while the current study considers thin films.
	To compare results, it is necessary to first estimate the strain in the \ce{V3Si} layers at the critical temperature, which can be done by extrapolating from the strain measured at room-temperature.
	
	For those crystals that have one of the cubic axes oriented vertically, we assume that the measured deviations in the room-temperature out-of-plane lattice parameters shown in Fig.~\ref{fig:XRDPlotStrainInsetTECPlusRT} correspond to such a tetragonal deformation.
	Their values are ${\epsilon_1=\SI{+0.2\pm0.4E-3}{}}$ and ${\epsilon_1=\SI{-3.3\pm0.1E-3}{}}$ on sapphire and silicon, respectively.
	The thermal contractions of the substrate materials and \ce{V3Si} down to cryogenic temperatures have been determined previously~\cite{wachtman1962linear,lyon1977linear,testardi1972structural,smith1975superconductivity}.
	Taking these into account as the samples cool from room temperature to the critical temperature of \ce{V3Si}, listed in Table~\ref{tab:induced_contractions}, we find the following estimates for the strain at \Tc:	
	\begin{equation}\epsilon_1^\text{sapphire}=\SI{-0.9\pm0.5E-3}{},\quad\epsilon_1^\text{silicon}=\SI{-5.2\pm0.2E-3}{}.\end{equation}
	When used in eq.~\eqref{eq:testardi_simplified}, these values correspond to reductions in the critical temperature between 0 and \SI{0.3}{\kelvin} on sapphire, and between 3.5 and \SI{4.2}{\kelvin} on silicon, compared to unstrained bulk \ce{V3Si}.
	Conversely, a reduction in \Tc\ of \SI{1.9\pm0.3}{\kelvin} would in the case of a purely tetragonal deformation be associated with an out-of-plane strain of \SI{3.7\pm0.3E-3}{}.
	The behavior of the polycrystalline layer during cooling is more complex, however, than a uniform strained contraction, since not all grains have one of the cubic axes aligned vertically.
	Differently oriented grains could exhibit different modes of deformation in response to an in-plane tension, and the picture is complicated by an unpredictable degree of tetragonal deformation prevalent in A-15 superconductors.

	\begin{table}%
		\centering%
		\caption{\label{tab:induced_contractions}The estimated contractions of sapphire~\cite{wachtman1962linear}, silicon~\cite{lyon1977linear} and \ce{V3Si}~\cite{testardi1972structural,smith1975superconductivity} from 300 to \SI{16}{\kelvin} (valid both in-plane and out-of-plane), and the induced out-of-plane expansion of \ce{V3Si}.}%
		\begin{tabular}{l@{\;\;\;}l@{\;\;\;}l@{\;\;\;}l}
			\hline
			Material	& Contraction 				& Relative to \ce{V3Si}	& Induced $\Delta\epsilon_1$\\\hline\\[-0.8em]
			Sapphire
						& $-\SI{6.2\pm0.8E-4}{}$	& $+\SI{5.8\pm1.3E-4}{}$& $-\SI{1.2\pm0.3E-3}{}$\\
			Silicon
						& $-\SI{2.33E-4}{}$			& $+\SI{9.7\pm1E-4}{}$	& $-\SI{1.9\pm0.2E-3}{}$\\
			\ce{V3Si}
						& $-\SI{1.2\pm0.1E-3}{}$	& ---					& ---\\\hline
		\end{tabular}
	\end{table}
	
	A Martensitic tetragonal deformation is observed in most, though not all, monocrystalline samples of \ce{V3Si}~\cite{batterman1964crystal,batterman1966low,testardi1970unusual}, and is associated with a strain up to ${\epsilon_1=-2\,\epsilon_{2,3}=\SI{1.7E-3}{}}$, where the unit cell contracts along two axes while it expands along the other, generally referred to as the $a=b$ and $c$ axes.
	It should be noted that those samples where the transformation did not occur were found to have lower resistance ratios ${R_{\SI{300}{\kelvin}}/R_{\SI{20.5}{\kelvin}}}$ than  those where it did~\cite{batterman1966low}.
	It is likely that this deformation occurs to some degree in strained polycrystalline samples as well, with the axes of expansion necessarily misaligned.
	
	The larger reduction in \Tc\ on silicon is due to an in-plane tensile strain.
	Therefore, grains where the expanding $c$-axis of the Martensitic transition is close to in-plane, the critical temperature would be further reduced.
	In others, the Martensitic deformation and the strain imposed by the substrate could partially compensate each other, where the compensation would be maximal when the $c$-axis points out of plane.
	The smaller relative reduction in \Tc\ on silicon than would be expected from the extrapolation of the measured room-temperature out-of-plane lattice parameters suggests that for a large share of the grains the Martensitic deformation compensates for the thermal stress.
	This implies that the substrate-induced strain selects a preferential axis for the Martensitic transition, such that the total deformation from a cubic crystal is minimized.	
	The dependence of the critical temperature on the orientation of the grains also provides a mechanism for the observed broadening of the superconducting transitions on the silicon substrate, as discussed earlier in section~\ref{sec:low_temperature}.
	
	Grains that are tilted relative to the vertical axis cannot accommodate the imposed strain with only a tetragonal deformation, and need to undergo a combination of volume change, plastic deformation and shearing.
	The first two are unlikely to contribute much, since \ce{V3Si} tends to preserve its unit cell volume~\cite{batterman1966low}, and the reversibility of the tetragonal deformation~\cite{batterman1964crystal} suggests that no plastic changes should occur.
	Shearing is more likely to take place considering the strong reduction in the restoring force for shear deformation at low temperatures~\cite{testardi1970unusual}.
	However, since~\cite{testardi1970unusual} ${\Delta_{44}\ll\Delta_{11}-\Delta_{12}}$, this is expected to contribute less to the broadening of the superconducting transitions than the misalignment of the $c$-axes of the Martensitic deformations of different grains in the polycrystalline film.
	
\section{Conclusion}
	
	Vanadium silicide is found to have a lower superconducting critical temperature on silicon substrates than it does on sapphire, a difference that cannot be straightforwardly ascribed to improvements in purity or crystallinity.
	In-situ XRD analysis during heating and cooling of as-deposited samples revealed that the thin films respond strongly to thermal expansion mismatches with the substrate, leaving the material relatively strained on silicon when cooled down after crystallization.
	This imposed strain is increased upon cooling to cryogenic temperatures, and is assumed to depend strongly on grain orientation due to the near vanishing of the shearing restoration force at low temperatures and the expected presence of a Martensitic transition just above \Tc.
	The observed reduction in critical temperature, as well as the broadening of the superconducting transitions, due to this inhomogeneous strain is well explained by an existing thermodynamic model for the A-15 superconducting compounds.
	
	Strain thus offers a pathway to critical temperature modulation of this class of superconductors.
	It may be relaxed by prolonged heating, in which case it is predicted to increase the critical temperature of \ce{V3Si} and \ce{V3Ga} thin films, and decrease that of \ce{V3Ge}~\cite{testardi1970unusual}.
	The engineering of local strain has previously been demonstrated on the scale of a few tens of nanometers~\cite{thompson200290,ghani200390nm}, an approach that could also be used to compensate for the stresses introduced by silicidation~\cite{steegen2002silicide}, or to spatially vary the superconducting properties of these materials.

\section{Acknowledgments}
	
	T.D.V. acknowledges the European Union’s Horizon 2020 research and innovation programme under the Marie Skłodowska-Curie grant agreement No 754303. This work was partially supported by the ANR project SUNISiDEUP (ANR-19-CE47-0010). JX Nippon are gratefully acknowledged for providing the \ce{V3Si} deposition target.

\bibliography{bibliography}

\end{document}